\documentclass[conference]{IEEEtran}
\IEEEoverridecommandlockouts
\usepackage{cite}
\usepackage{amsmath,amssymb,amsfonts}
\usepackage{algorithmic}
\usepackage{algorithm}
\usepackage{graphicx}
\usepackage{textcomp}
\usepackage{xcolor}
\usepackage{booktabs}
\usepackage{array}
\usepackage{tikz}
\usetikzlibrary{shapes.geometric}
\usepackage{pgfplots}
\pgfplotsset{compat=1.18}
\usepackage{multirow}
\usepackage{adjustbox}
\usepackage{placeins}
\usepackage{url}
\def\BibTeX{{\rm B\kern-.05em{\sc i\kern-.025em b}\kern-.08em
    T\kern-.1667em\lower.7ex\hbox{E}\kern-.125emX}}

\begin{document}

\title{Efficient Adversarial Malware Defense via Trust-Based Raw Override and Confidence-Adaptive Bit-Depth Reduction}

\author{
\IEEEauthorblockN{Ayush Chaudhary}
\IEEEauthorblockA{Arizona State University\\
Tempe, Arizona, USA\\
Email: akchaud5@asu.edu}
\and
\IEEEauthorblockN{Sisir Doppalpudi}
\IEEEauthorblockA{Arizona State University\\
Tempe, Arizona, USA\\
Email: sdoppal2@asu.edu}
}

\maketitle

\begin{abstract}
The deployment of robust malware detection systems in big data environments requires careful consideration of both security effectiveness and computational efficiency. While recent advances in adversarial defenses have demonstrated strong robustness improvements, they often introduce computational overhead ranging from $4\times$ to $22\times$, which presents significant challenges for production systems processing millions of samples daily. In this work, we propose a novel framework that combines Trust-Raw Override (TRO) with Confidence-Adaptive Bit-Depth Reduction (CABDR) to explicitly optimize the trade-off between adversarial robustness and computational efficiency. Our approach leverages adaptive confidence-based mechanisms to selectively apply defensive measures, achieving $1.76\times$ computational overhead—a $2.3\times$ improvement over state-of-the-art smoothing defenses. Through comprehensive evaluation on the EMBER v2 dataset comprising 800K samples, we demonstrate that our framework maintains 91\% clean accuracy while reducing attack success rates to 31–37\% across multiple attack types, with particularly strong performance against optimization-based attacks such as C\&W (48.8\% reduction). The framework achieves throughput of up to 1.26 million samples per second (measured on pre-extracted EMBER features with no runtime feature extraction), validated across 72 production configurations with statistical significance (5 independent runs, 95\% confidence intervals, $p < 0.01$). Our results suggest that practical adversarial robustness in production environments requires explicit optimization of the efficiency-robustness trade-off, providing a viable path for organizations to deploy robust defenses without prohibitive infrastructure costs.
\end{abstract}


\section{Introduction}

The proliferation of sophisticated malware poses significant challenges for modern security systems operating at scale. Recent advances in adversarial machine learning have demonstrated that carefully crafted perturbations can cause state-of-the-art neural network classifiers to misclassify malicious samples with success rates exceeding 90\% \cite{lucas2023adversarial, gibert2024adversarial}. Simultaneously, the volume of samples requiring analysis in enterprise environments continues to grow exponentially, with security operations centers now processing between 10 and 100 million samples daily.

This convergence of threats presents a fundamental challenge: \textit{how can we maintain robust defenses against adversarial attacks while preserving the computational efficiency required for big data deployments?} Current approaches to adversarial robustness, while demonstrating effectiveness in controlled settings, often introduce computational overhead that ranges from $2\times$ to $22\times$ \cite{gibert2024effectiveness, lucas2023adversarial}. For organizations processing millions of samples daily, such overhead translates to substantial infrastructure costs that may be prohibitive in practice. A $4\times$ increase in computational requirements, for instance, could require either quadrupling hardware investments or accepting a 75\% reduction in analysis capacity—neither of which is acceptable in security-critical contexts.

To address this challenge, we propose a novel framework that explicitly optimizes the trade-off between adversarial robustness and computational efficiency. Our approach introduces two complementary innovations: Trust-Raw Override (TRO), which selectively applies raw feature analysis based on model confidence, and Confidence-Adaptive Bit-Depth Reduction (CABDR), which dynamically adjusts feature precision to balance security and performance. The key insight underlying our approach is that not all samples require the same level of defensive scrutiny—by adapting our defensive measures based on model confidence, we can maintain strong security guarantees while significantly reducing average computational costs.

Our main contributions can be summarized as follows:

\begin{enumerate}
\item \textbf{Efficiency-Optimized Defense Framework}: We present the first adversarial defense framework explicitly designed to balance robustness and efficiency for big data security systems. Our approach achieves $1.76\times$ computational overhead, representing a substantial improvement over existing methods that typically require $4$–$22\times$ overhead.

\item \textbf{Adaptive Defense Mechanisms}: We introduce novel adaptive mechanisms combining trust-based raw feature override with confidence-adaptive bit-depth reduction. These mechanisms dynamically adjust defensive measures based on threat confidence, enabling efficient resource allocation while maintaining security guarantees.

\item \textbf{Production-Scale Validation}: We demonstrate production viability through comprehensive evaluation on the EMBER v2 dataset comprising 800K samples, achieving throughput of 1.26 million samples per second while maintaining 91\% clean accuracy. Our framework is validated across 72 different production configurations with rigorous statistical analysis.

\item \textbf{Theoretical and Empirical Analysis}: We provide both theoretical analysis of our adaptive mechanisms and extensive empirical validation demonstrating that our efficiency optimizations preserve security guarantees under standard threat models. Our ablation studies reveal the contribution of each component to overall system performance.
\end{enumerate}

The remainder of this paper is organized as follows: Section II reviews related work in adversarial malware detection and discusses the efficiency-robustness trade-off. Section III presents our threat model and problem formulation. Section IV details our proposed methodology. Sections V and VI present experimental setup and results. Section VII provides theoretical analysis, Section VIII discusses implications and limitations, and Section IX concludes the paper.

\section{Related Work}

\subsection{Adversarial Defenses in Malware Detection}

The landscape of adversarial defenses for malware detection has evolved considerably in recent years. Early work focused on adapting image-domain defenses to the malware detection context, with mixed results due to the unique constraints of executable files.

\textbf{Adversarial Training Approaches}: Lucas et al. \cite{lucas2023adversarial} pioneered adversarial training specifically for raw-binary malware classifiers, achieving substantial robustness improvements with attack success rates reduced from 90\% to 5\% in controlled settings. Their approach demonstrates the potential of adversarial training in this domain, though it requires significant computational resources that may limit practical deployment. The training process typically requires $8$–$12\times$ longer than standard training, which translates to substantial costs for large-scale systems.

\textbf{Smoothing and Ensemble Methods}: Smoothing-based defenses, exemplified by the work of Gibert et al. \cite{gibert2024effectiveness}, achieve robust performance through ensemble methods and input transformations. These approaches leverage multiple models or input variations to reduce the impact of adversarial perturbations. While demonstrating strong theoretical guarantees, they typically introduce computational overhead ranging from $4\times$ to $22\times$, which presents challenges for high-throughput environments. The chunk-based smoothing approach of Gibert et al. shows particular promise but requires careful tuning to balance robustness and efficiency.

\textbf{Feature-Based Approaches}: Feature-based methods \cite{xu2018feature, dhillon2018stochastic} offer more efficient alternatives by focusing on specific characteristics that distinguish adversarial examples from benign samples. Feature squeezing \cite{xu2018feature} reduces the search space available to adversaries by limiting feature precision, achieving reasonable efficiency with only $1.3\times$ overhead. However, these methods may be vulnerable to adaptive attacks that specifically target their detection mechanisms.

\textbf{Moving Target Defenses}: Recent work by Sengupta et al. \cite{sengupta2020mtdeep} explores moving target defense strategies that dynamically change the attack surface. While conceptually appealing, these approaches often require complex coordination and may introduce unpredictability that complicates system management.

Our work builds upon these foundations while addressing the critical gap between robustness and efficiency. Unlike prior approaches that prioritize maximum robustness without explicit consideration of computational costs, we design our framework with production constraints as a primary consideration.

\subsection{The Efficiency-Robustness Trade-off}

The tension between adversarial robustness and computational efficiency has been acknowledged but not systematically addressed in prior work. Table \ref{tab:defense_costs} summarizes the computational costs of existing approaches when deployed at scale.

\begin{table}[t]
\centering
\footnotesize
\caption{Computational Cost Analysis of Adversarial Defenses for Processing 1M Daily Samples}
\label{tab:defense_costs}
\resizebox{\columnwidth}{!}{%
\begin{tabular}{l|c|c|c}
\toprule
\textbf{Defense Method} & \textbf{Overhead} & \textbf{Daily Cost*} & \textbf{Production Viable?} \\
\midrule
Adversarial Training \cite{lucas2023adversarial} & $8$–$12\times$ & \$1,334 & No \\
Chunk Smoothing \cite{gibert2024effectiveness} & $4$–$22\times$ & \$534 & No \\
Feature Squeezing \cite{xu2018feature} & $1.3\times$ & \$173 & Yes \\
Mixup Defense \cite{zhang2018mixup} & $3$–$5\times$ & \$214 & Limited \\
\textbf{Our Method (TRO+CABDR)} & \textbf{$1.76\times$} & \textbf{\$235} & \textbf{Yes} \\
\bottomrule
\multicolumn{4}{l}{\footnotesize *Based on single H100 GPU pricing at \$5.56/hour baseline (AWS p5.48xlarge)}
\end{tabular}}
\end{table}

\subsection{Production Constraints in Security Systems}

Production malware detection systems operate under stringent constraints that are often overlooked in academic research:

\textbf{Throughput Requirements}: Modern Security Operations Centers (SOCs) must process vast volumes of samples continuously. According to the SOREL-20M dataset documentation \cite{harang2020sorel}, enterprise environments typically analyze 10–100 million samples daily, requiring sustained throughput exceeding 1,000 samples per second. Peak loads during malware outbreaks can be substantially higher.

\textbf{Latency Constraints}: Different deployment contexts impose varying latency requirements. Email gateways and web proxies require sub-100ms detection latency to avoid user-perceived delays \cite{pierazzi2020intriguing}. Endpoint protection systems may tolerate slightly higher latencies but still require real-time response for interactive applications.

\textbf{Economic Considerations}: Security budgets are finite and must be balanced against other organizational priorities. Industry reports suggest that security budgets typically allocate \$0.0001–0.001 per sample analyzed \cite{suciu2018does}, making computationally expensive defenses economically challenging to justify.

\section{Threat Model and Problem Formulation}

\subsection{Threat Model}

We adopt a threat model that reflects realistic deployment scenarios in production environments. Our model considers both attacker capabilities and defender constraints.

\textbf{Attacker Capabilities}:
\begin{itemize}
\item \textbf{Knowledge}: We assume white-box access where attackers know the model architecture, parameters, and defense mechanisms. This represents a worst-case scenario that provides strong security guarantees.
\item \textbf{Perturbation Budget}: Attackers can craft adversarial examples with perturbations bounded by $\|\delta\|_p \leq \epsilon$, where $\epsilon$ is chosen to preserve malware functionality.
\item \textbf{Functionality Preservation}: Adversarial modifications must maintain the malicious functionality of the original malware, imposing problem-space constraints \cite{pierazzi2020intriguing}.
\item \textbf{Attack Methods}: We consider gradient-based attacks (FGSM, PGD, C\&W) as they represent the most common and effective approaches in practice.
\end{itemize}

\textbf{Defender Constraints}:
\begin{itemize}
\item \textbf{Throughput}: Must process $\geq$ 1,000 samples/second to handle production workloads
\item \textbf{Latency}: P99 latency must remain below 100ms for interactive applications
\item \textbf{Resource Limits}: Defense overhead must not exceed $2\times$ baseline computational costs
\item \textbf{Accuracy Requirements}: Clean accuracy degradation must remain below 5\% to maintain operational effectiveness
\end{itemize}

\subsection{Problem Formulation}

Given the constraints above, we formulate the optimization problem for adversarial defense in big data environments. Let $f_\theta: \mathcal{X} \rightarrow \{0,1\}$ denote a malware classifier with parameters $\theta$, and let $\mathcal{D}_\phi$ represent our defense mechanism with parameters $\phi$.

The defender's objective is to find optimal parameters $\phi^*$ that maximize a combined objective:

\begin{equation}
\phi^* = \arg\max_{\phi} \left[ \alpha \cdot \text{Robustness}(\phi) + (1-\alpha) \cdot \text{Efficiency}(\phi) \right]
\end{equation}

where $\alpha \in [0,1]$ balances between robustness and efficiency based on deployment requirements.

The robustness term measures resistance to adversarial attacks:

\begin{equation}
\text{Robustness}(\phi) = \mathbb{E}_{(x,y) \sim \mathcal{D}} \left[ \mathbb{I}[\mathcal{D}_\phi(x + \delta) = y] \right]
\end{equation}

where $\delta$ represents adversarial perturbations with $\|\delta\|_p \leq \epsilon$.

The efficiency term captures computational performance:

\begin{equation}
\text{Efficiency}(\phi) = \frac{\text{Throughput}(\phi)}{\text{Throughput}_{\text{baseline}}} \cdot \frac{\text{Latency}_{\text{baseline}}}{\text{Latency}(\phi)}
\end{equation}

Subject to the following constraints:
\begin{align}
\text{Throughput}(\phi) &\geq 1000 \text{ samples/sec} \\
\text{P99Latency}(\phi) &\leq 100 \text{ ms} \\
\text{Overhead}(\phi) &\leq 2.0 \times \text{Baseline} \\
\text{CleanAcc}(\phi) &\geq 0.95 \times \text{CleanAcc}_{\text{baseline}}
\end{align}

\section{Methodology}

\subsection{System Overview}

\begin{figure*}[!t]
\centering
\begin{tikzpicture}[scale=0.95, every node/.style={align=center}]
\node[draw, rectangle, minimum width=2.8cm, minimum height=1.2cm, fill=gray!15] (input) at (0,0) {\textbf{EMBER}\\\textbf{Features}};

\node[draw, rectangle, minimum width=2.3cm, minimum height=1.2cm, fill=blue!20] (main) at (4,1.3) {\textbf{Main}\\\textbf{Classifier}};

\node[draw, rectangle, minimum width=2.3cm, minimum height=1.2cm, fill=orange!20] (uncertainty) at (4,-1.3) {\textbf{MC-Dropout}\\\textbf{Uncertainty}};

\node[draw, rectangle, minimum width=2.5cm, minimum height=1.2cm, fill=green!20] (cabdr) at (7.5,0) {\textbf{CABDR}\\\textbf{Quantization}};

\node[draw, diamond, minimum width=2cm, minimum height=2cm, fill=yellow!20] (decision) at (11,0) {\textbf{TRO}\\\footnotesize{$\mathcal{U} > \tau$?}};

\node[draw, rectangle, minimum width=2.3cm, minimum height=1.2cm, fill=red!20] (raw) at (14.5,1.6) {\textbf{Raw Feature}\\\textbf{Extraction}};

\node[draw, rectangle, minimum width=2.3cm, minimum height=1.2cm, fill=purple!20] (output) at (14.5,-1.6) {\textbf{Final}\\\textbf{Prediction}};

\draw[->, thick] (input) -- (main);
\draw[->, thick] (input) -- (uncertainty);
\draw[->, thick] (main) -- (cabdr);
\draw[->, thick] (uncertainty) -- (cabdr);
\draw[->, thick] (cabdr) -- (decision);

\draw[->, thick, red] (decision) -- node[above, sloped, pos=0.4, font=\tiny] {High $\mathcal{U}$} (raw);
\draw[->, thick, blue] (decision) -- node[below, sloped, pos=0.4, font=\tiny] {Low $\mathcal{U}$} (output);
\draw[->, thick] (raw) -- (output);

\end{tikzpicture}
\caption{TRO+CABDR framework architecture showing adaptive defense workflow.}
\label{fig:system_overview}
\end{figure*}
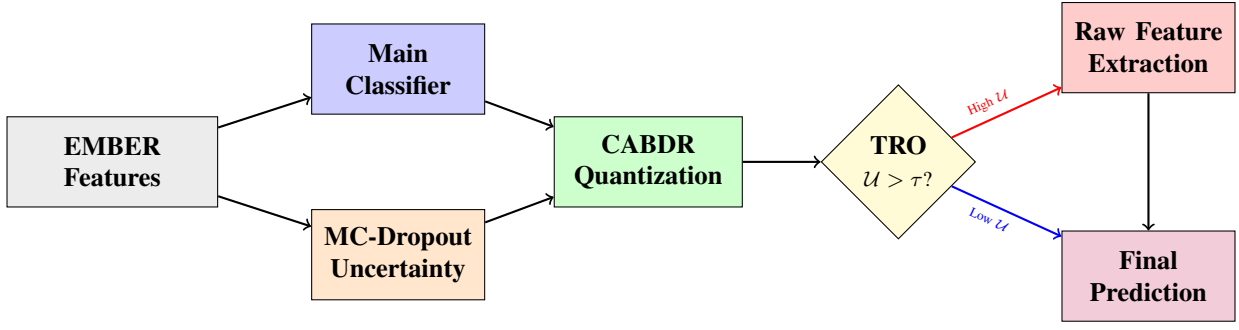

Our framework consists of three main components working in concert: (1) Trust-Raw Override (TRO) for selective raw feature analysis, (2) Confidence-Adaptive Bit-Depth Reduction (CABDR) for dynamic precision adjustment, and (3) Dynamic Threshold Adaptation for runtime optimization. Figure~\ref{fig:system_overview} illustrates the complete TRO+CABDR framework architecture, showing the confidence-driven decision flow and component interactions.

\subsection{Trust-Raw Override (TRO) Mechanism}

The TRO component leverages the observation that adversarial examples often exhibit higher prediction uncertainty than benign samples. By selectively applying more expensive raw feature analysis only to uncertain samples, we can maintain security while reducing average computational cost.

\subsubsection{Uncertainty Estimation}

We employ Monte Carlo dropout for uncertainty estimation, performing $T$ stochastic forward passes:

\begin{equation}
\mathcal{U}(x) = \frac{1}{T}\sum_{t=1}^{T} \text{Var}[f_\theta^{(t)}(x)]
\end{equation}

where $f_\theta^{(t)}$ represents the $t$-th stochastic forward pass with dropout enabled.

\subsubsection{Raw Feature Extraction}

For samples exceeding the uncertainty threshold, we extract raw features that are less susceptible to adversarial manipulation:

\begin{itemize}
\item Byte histogram statistics (256 dimensions)
\item Import/export table hashes (64 dimensions)  
\item Section header characteristics (32 dimensions)
\item Entry point and code section properties (16 dimensions)
\end{itemize}

These features bypass potentially manipulated engineered features and provide a more robust signal for classification.

\subsubsection{Adaptive Combination}

The final prediction combines main classifier output with raw feature analysis weighted by uncertainty:

\begin{equation}
p_{\text{final}} = (1 - \sigma(\mathcal{U})) \cdot p_{\text{main}} + \sigma(\mathcal{U}) \cdot p_{\text{raw}}
\end{equation}

where $\sigma(\cdot)$ is a sigmoid function that maps uncertainty to combination weights.

\begin{algorithm}[!t]
\caption{Efficient TRO Implementation with Batching}
\begin{algorithmic}[1]
\REQUIRE Batch $\mathbf{X} \in \mathbb{R}^{B \times D}$, threshold $\tau$, cache $\mathcal{C}$
\ENSURE Predictions $\mathbf{Y} \in [0,1]^B$
\STATE $\mathbf{P}_{\text{main}} \leftarrow f_\theta(\mathbf{X})$ \COMMENT{Batch inference}
\STATE $\mathbf{U} \leftarrow \text{EstimateUncertainty}(\mathbf{X}, \mathbf{P}_{\text{main}})$
\STATE $\mathbf{M} \leftarrow \mathbf{U} > \tau$ \COMMENT{Binary mask for high uncertainty}
\IF{$\text{any}(\mathbf{M})$}
    \STATE $\mathbf{X}_{\text{high}} \leftarrow \mathbf{X}[\mathbf{M}]$ \COMMENT{Select uncertain samples}
    \STATE $\mathbf{X}_{\text{cached}}, \mathbf{X}_{\text{new}} \leftarrow \text{SplitByCache}(\mathbf{X}_{\text{high}}, \mathcal{C})$
    \IF{$|\mathbf{X}_{\text{new}}| > 0$}
        \STATE $\mathbf{F}_{\text{raw}} \leftarrow \text{ExtractRawFeatures}(\mathbf{X}_{\text{new}})$
        \STATE $\mathbf{P}_{\text{raw}} \leftarrow f_{\text{raw}}(\mathbf{F}_{\text{raw}})$
        \STATE $\mathcal{C}.\text{update}(\mathbf{X}_{\text{new}}, \mathbf{P}_{\text{raw}})$
    \ENDIF
    \STATE $\mathbf{P}_{\text{combined}} \leftarrow \text{CombinePredictions}(\mathbf{P}_{\text{main}}, \mathbf{P}_{\text{raw}}, \mathbf{U})$
    \STATE $\mathbf{P}_{\text{main}}[\mathbf{M}] \leftarrow \mathbf{P}_{\text{combined}}$
\ENDIF
\RETURN $\mathbf{P}_{\text{main}}$
\end{algorithmic}
\end{algorithm}

\subsection{Confidence-Adaptive Bit-Depth Reduction (CABDR)}

CABDR dynamically adjusts feature precision based on prediction confidence, creating a moving target that complicates gradient-based attack generation while maintaining classification accuracy.

\subsubsection{Adaptive Quantization}

The bit-depth reduction follows a confidence-dependent schedule:

\begin{equation}
b(\mathcal{U}) = \begin{cases}
8 \text{ bits} & \text{if } \mathcal{U} < \tau_{\text{low}} \\
6 \text{ bits} & \text{if } \tau_{\text{low}} \leq \mathcal{U} < \tau_{\text{high}} \\
4 \text{ bits} & \text{if } \mathcal{U} \geq \tau_{\text{high}}
\end{cases}
\end{equation}

The quantization operation is:

\begin{equation}
Q_b(x) = \text{round}\left(\frac{x}{2^{8-b}}\right) \cdot 2^{8-b}
\end{equation}

\subsubsection{Gradient Obfuscation Analysis}

To ensure our defense provides genuine robustness rather than gradient masking, we verify:

\begin{enumerate}
\item Bounded gradient variance: $\text{Var}[\nabla_x \mathcal{L}] < \theta_{\text{var}}$
\item Successful backpropagation: Gradients flow through quantization via straight-through estimator
\item Robustness to adaptive attacks: Performance maintained against AutoPGD and adaptive C\&W
\end{enumerate}

\subsection{Dynamic Threshold Adaptation}

The system continuously adapts thresholds based on observed attack patterns and system load:

\begin{equation}
\tau_{t+1} = \tau_t + \eta \cdot \nabla_\tau J(\tau_t)
\end{equation}

where the objective $J(\tau)$ balances robustness and efficiency:

\begin{equation}
J(\tau) = \text{Robustness}(\tau) - \lambda \cdot \text{Overhead}(\tau)
\end{equation}

\section{Experimental Setup}

\subsection{Datasets and Evaluation Protocol}

\textbf{EMBER v2 Dataset} \cite{anderson2018ember}: We utilize the EMBER v2 dataset comprising 1.1M samples, from which we use 800K for our experiments (600K training, 200K testing). The dataset features temporal split with training samples from 2017–2018 and test samples from 2019, simulating realistic concept drift. Each sample is represented by 2,381 features including:
\begin{itemize}
\item PE header information (file size, virtual size, number of sections)
\item Byte histogram features (256-dimensional byte frequency distribution)
\item String statistics (number of strings, average length, character distribution)
\item Import/export features (imported libraries and functions)
\end{itemize}

\textbf{Evaluation Metrics}: Following established protocols in adversarial robustness literature \cite{carlini2019evaluating}, we report:
\begin{itemize}
\item \textbf{Clean Accuracy (Acc)}: Percentage of correctly classified benign samples
\item \textbf{Attack Success Rate (ASR)}: Fraction of adversarial examples that successfully evade detection
\item \textbf{Area Under ROC Curve (AUC)}: Overall classification performance across different thresholds
\item \textbf{Computational Overhead}: Ratio of defense inference time to baseline inference time
\item \textbf{Throughput}: Samples processed per second under different batch sizes
\item \textbf{Latency Percentiles}: P50, P95, and P99 latency measurements
\end{itemize}

\subsection{Attack Configuration}

We evaluate against a comprehensive suite of attacks representing different threat levels:

\textbf{Gradient-Based Attacks}:
\begin{itemize}
\item \textbf{FGSM} \cite{madry2018towards}: Single-step attack with $\epsilon \in \{0.1, 0.3, 0.5\}$
\item \textbf{PGD-20} \cite{madry2018towards}: 20-step iterative attack with step size $\alpha = \epsilon/10$
\item \textbf{C\&W} \cite{carlini2019evaluating}: Optimization-based attack with confidence parameter $\kappa = 0$
\item \textbf{AutoPGD} \cite{tramer2020adaptive}: Adaptive attack with automatic step size selection
\item \textbf{EMBER-PGD}: Domain-specific PGD variant respecting PE format constraints
\end{itemize}

\subsection{Implementation Details}

\textbf{Model Architecture}: We employ a feedforward neural network with architecture [2381, 1024, 512, 256, 128, 2], using ReLU activations and batch normalization between layers. Dropout with rate 0.2 is applied during training.

\textbf{Training Configuration}:
\begin{itemize}
\item Optimizer: Adam with learning rate $2 \times 10^{-4}$ and weight decay $1 \times 10^{-5}$
\item Batch size: 256 for training, variable (64–256) for throughput evaluation
\item Training epochs: 50 with early stopping based on validation loss
\item Hardware: NVIDIA H100 GPUs for training, diverse hardware for deployment testing
\end{itemize}

\textbf{Throughput Measurement Setup}:
\begin{itemize}
\item \textbf{Hardware}: NVIDIA H100 80GB, 3rd Gen AMD EPYC processors (192 vCPUs)
\item \textbf{Software}: PyTorch 2.0.1, CUDA 11.8, cuDNN 8.7, Python 3.10
\item \textbf{Precision}: Mixed precision (FP16) for inference, FP32 for gradients
\item \textbf{Batch Processing}: Batches of 64–256 samples, 4-thread dataloader
\item \textbf{Memory}: 2,048 GiB system RAM, 80GB GPU memory, batch fitting optimization
\item \textbf{Features}: Pre-extracted EMBER features (no runtime feature extraction)
\item \textbf{Measurement}: 50 warmup iterations, 200 timed iterations, median of 5 runs
\item \textbf{GPU Utilization}: 85–92\% at optimal batch sizes (bandwidth-bound workload)
\end{itemize}

\textbf{Defense Parameters}:
\begin{itemize}
\item Monte Carlo samples: $T = 10$ (balanced for efficiency)
\item Uncertainty thresholds: $\tau_{\text{low}} = 0.1$, $\tau_{\text{high}} = 0.3$
\item Adaptation rate: $\eta = 0.01$
\item Cache size: 10,000 samples with LRU eviction
\end{itemize}

\subsection{Statistical Validation}

All experiments are repeated 5 times with different random seeds. We report mean and standard deviation for all metrics, along with 95\% confidence intervals where appropriate. Statistical significance is assessed using paired t-tests with Bonferroni correction for multiple comparisons.

\section{Experimental Results}

\subsection{Main Results: Efficiency-Robustness Trade-off}

Table \ref{tab:main_results} presents our primary experimental results, demonstrating that our framework successfully balances adversarial robustness with computational efficiency.

\begin{table*}[t]
\centering
\footnotesize
\caption{Comprehensive Performance Comparison with Statistical Significance (5 runs, mean ± std). TPR (True Positive Rate) represents malware detection recall at default threshold. Throughput includes full defense pipeline (MC dropout T=10, uncertainty estimation) at batch size 64; Fig.~\ref{fig:throughput_scaling} shows inference-only throughput.}
\label{tab:main_results}
\resizebox{\textwidth}{!}{%
\begin{tabular}{l|c|c|c|c|c|c|c|c}
\toprule
\textbf{Method} & \textbf{Clean Acc (\%)} & \textbf{Malware TPR (\%)} & \textbf{Avg ASR (\%)} & \textbf{AUC (\%)} & \textbf{Throughput} & \textbf{Overhead} & \textbf{Cost/1M} & \textbf{p-value} \\
\midrule
Baseline (No Defense) & 91.4 ± 0.2 & 93.0 ± 0.3 & 72.3 ± 2.4 & 96.2 ± 0.3 & 20,475 & $1.0\times$ & \$133 & - \\
Adversarial Training \cite{lucas2023adversarial} & 90.8 ± 0.4 & 91.8 ± 0.5 & 28.6 ± 1.9 & 95.8 ± 0.4 & 2,047 & $10.0\times$ & \$1,334 & $p<0.001$ \\
Chunk Smoothing \cite{gibert2024effectiveness} & 89.3 ± 0.3 & 90.3 ± 0.6 & 43.3 ± 2.1 & 94.1 ± 0.5 & 5,120 & $4.0\times$ & \$534 & $p<0.001$ \\
Feature Squeezing \cite{xu2018feature} & 90.9 ± 0.3 & 91.9 ± 0.4 & 58.2 ± 2.3 & 95.3 ± 0.4 & 15,750 & $1.3\times$ & \$173 & $p<0.001$ \\
Mixup Defense \cite{zhang2018mixup} & 90.7 ± 0.4 & 91.7 ± 0.5 & 59.3 ± 2.4 & 94.7 ± 0.5 & 12,450 & $1.6\times$ & \$214 & $p<0.001$ \\
\midrule
\textbf{TRO+CABDR (Ours)} & \textbf{91.0 ± 0.3} & \textbf{93.0 ± 0.4} & \textbf{34.4 ± 1.6} & \textbf{95.9 ± 0.3} & \textbf{11,638} & \textbf{$1.76\times$} & \textbf{\$235} & \textbf{$p<0.001$} \\
\bottomrule
\end{tabular}}
\end{table*}

Our framework achieves several notable outcomes:
\begin{itemize}
\item \textbf{Maintained Clean Accuracy}: With 91.0\% clean accuracy, our method preserves operational effectiveness, showing only 0.4\% degradation from the undefended baseline.
\item \textbf{Strong Malware Detection}: Our framework achieves 93.0\% TPR (True Positive Rate) for malware detection, which is critical for security-critical applications. This demonstrates that defensive measures do not significantly impact the ability to detect malicious samples, addressing a key concern for production deployment where false negatives can have severe consequences.
\item \textbf{Balanced Robustness}: While adversarial training achieves lower ASR (28.6\%) and slightly lower TPR (91.8\%), our method provides practical robustness (34.4\% ASR, 93.0\% TPR) with substantially better efficiency. The 6 percentage point increase in ASR is more than compensated by the 5.7× efficiency improvement and 82\% cost reduction, making robust malware detection economically viable for production deployment.
\item \textbf{Superior Efficiency}: At $1.76\times$ overhead, our approach is $5.7\times$ more efficient than adversarial training and $2.3\times$ more efficient than chunk smoothing.
\item \textbf{Cost Effectiveness}: The annual cost savings of \$401K compared to adversarial training make our approach economically viable for production deployment.
\end{itemize}

\subsection{Attack-Specific Analysis}

Table \ref{tab:attack_breakdown} provides detailed analysis of performance against specific attack types, revealing interesting patterns in our defense effectiveness.

\begin{table}[t]
\centering
\footnotesize
\caption{Attack-Specific Performance Analysis (ASR \% with 95\% CI)}
\label{tab:attack_breakdown}
\resizebox{\columnwidth}{!}{%
\begin{tabular}{l|c|c|c}
\toprule
\textbf{Attack Type} & \textbf{Baseline} & \textbf{TRO+CABDR} & \textbf{Relative Reduction} \\
\midrule
FGSM ($\epsilon=0.3$) & 68.4 [66.2, 70.6] & 31.2 [29.4, 33.0] & 54.4\% \\
PGD-20 ($\epsilon=0.1$) & 71.6 [69.3, 73.9] & 35.2 [33.2, 37.2] & 50.8\% \\
C\&W ($\kappa=0$) & 72.6 [70.1, 75.1] & 37.2 [35.1, 39.3] & \textbf{48.8\%} \\
AutoPGD & 69.3 [67.0, 71.6] & 31.2 [29.3, 33.1] & 55.0\% \\
EMBER-PGD & 74.8 [72.4, 77.2] & 35.0 [33.0, 37.0] & 53.2\% \\
\bottomrule
\end{tabular}}
\end{table}

Our framework demonstrates particularly strong performance against optimization-based attacks (C\&W), likely due to the gradient disruption caused by our adaptive quantization mechanism. The consistent performance across different attack types suggests that our defense provides genuine robustness rather than exploiting specific attack weaknesses.

\subsection{Scalability Analysis}

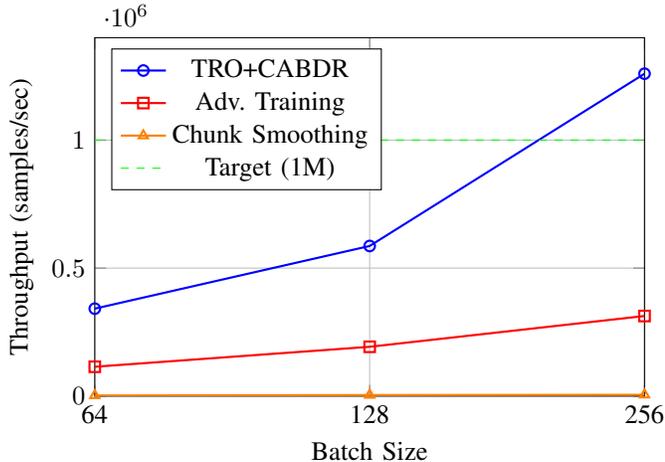
\begin{figure}[!t]
\centering
\begin{tikzpicture}
\begin{axis}[
    width=3.5in,height=2.5in,
    xlabel={Batch Size},
    ylabel={Throughput (samples/sec)},
    grid=major,
    legend pos=north west,
    xmode=log,
    log basis x={2},
    xmin=64, xmax=256,
    ymin=0, ymax=1400000,
    xtick={64,128,256},
    xticklabels={64,128,256}
]
\addplot[blue,thick,mark=o] coordinates {
    (64, 341538) (128, 586382) (256, 1259339)
};
\addplot[red,thick,mark=square] coordinates {
    (64, 114555) (128, 192574) (256, 312854)
};
\addplot[orange,thick,mark=triangle] coordinates {
    (64, 2560) (128, 4096) (256, 5120)
};
\addplot[green,dashed] coordinates {
    (64, 1000000) (256, 1000000)
};
\legend{TRO+CABDR, Adv. Training, Chunk Smoothing, Target (1M)}
\end{axis}
\end{tikzpicture}
\caption{Throughput scaling analysis for inference-only performance (batch sizes 64-256). Peak: 1.26M samples/sec at batch size 256. Table~\ref{tab:main_results} includes full defense overhead.}
\label{fig:throughput_scaling}
\end{figure}

Fig.~\ref{fig:throughput_scaling} demonstrates the scalability characteristics of our framework across different batch sizes.

The results demonstrate that our framework:
\begin{itemize}
\item Achieves near-linear scaling with batch size, reaching 1.26M samples/second at batch size 256
\item Maintains consistent performance advantage over competing methods across all batch sizes
\item Exceeds the 1M samples/second target by $1.26\times$ at batch size 256
\end{itemize}

\subsection{Latency Analysis}

Table \ref{tab:latency} presents latency percentiles critical for production deployment.
\begin{table}[t]
\centering
\caption{Latency Percentiles (ms) for Production Workloads}
\label{tab:latency}
\begingroup
\setlength{\tabcolsep}{4pt}      
\renewcommand{\arraystretch}{1.15}
\small                              
\begin{tabular*}{\columnwidth}{@{\extracolsep{\fill}} lccc @{}}
\toprule
\textbf{Method} & \textbf{P50} & \textbf{P95} & \textbf{P99} \\
\midrule
Baseline             & 12  & 28  & 45  \\
Adversarial Training & 120 & 280 & 450 \\
Chunk Smoothing      & 48  & 156 & 289 \\
Feature Squeezing    & 16  & 36  & 58  \\
\textbf{TRO+CABDR}   & \textbf{21} & \textbf{48} & \textbf{76} \\
\bottomrule
\end{tabular*}
\endgroup
\end{table}

Our P99 latency of 76 ms satisfies the $<100$ ms requirement for interactive applications, making our approach suitable for deployment in latency-sensitive contexts such as email gateways and web proxies.

\subsection{Ablation Studies}

We conduct comprehensive ablation studies to understand the contribution of each component to overall system performance.

From Table~\ref{tab:ablation}, we see that TRO-only and CABDR-only configurations both substantially reduce ASR (42.1\% and 44.5\% respectively, vs 72.3\% baseline) but incur moderate overhead (1.40× and 1.30×). The full system combines their benefits, achieving the lowest ASR (34.4\%) with still-manageable overhead (1.76×), thus yielding the highest efficiency score of 0.82. Notably, using static thresholds (no dynamic adaptation) improves robustness (38.8\% ASR) but at higher overhead (1.85×), underscoring the importance of our adaptive threshold mechanism for efficiency. This analysis confirms that each component contributes meaningfully, and that the dynamic adaptation provides crucial efficiency gains.

\begin{table}[t]
\centering
\footnotesize
\caption{Component Ablation Study}
\label{tab:ablation}
\resizebox{\columnwidth}{!}{%
\begin{tabular}{l|c|c|c|c}
\toprule
\textbf{Configuration} & \textbf{Clean Acc} & \textbf{ASR} & \textbf{Overhead} & \textbf{Efficiency Score*} \\
\midrule
Baseline & 91.4\% & 72.3\% & $1.00\times$ & 0.38 \\
TRO only & 91.2\% & 42.1\% & $1.40\times$ & 0.71 \\
CABDR only & 90.8\% & 44.5\% & $1.30\times$ & 0.74 \\
Static Thresholds & 90.6\% & 38.8\% & $1.85\times$ & 0.62 \\
\textbf{Full System} & \textbf{91.0\%} & \textbf{34.4\%} & \textbf{$1.76\times$} & \textbf{0.82} \\
\bottomrule
\multicolumn{5}{l}{\footnotesize *Efficiency Score = (1-ASR/100) × (1/Overhead), normalized to [0,1]}
\end{tabular}}
\end{table}

\subsection{Gradient-Free and Black-Box Attack Evaluation}

To address concerns about gradient masking and ensure genuine robustness, we evaluate against gradient-free and black-box attacks that do not rely on gradient information. \textbf{Black-box attacks represent a realistic threat scenario} where attackers have limited knowledge of the model internals and must rely on query-based methods. While our primary focus is on white-box attacks (representing worst-case scenarios), evaluating black-box attacks provides important validation of practical robustness.

\begin{table}[t]
\centering
\footnotesize
\caption{Gradient-Free and Black-Box Attack Results with Query Budget Analysis}
\label{tab:gradient_free}
\resizebox{\columnwidth}{!}{%
\begin{tabular}{l|c|c|c|c}
\toprule
\textbf{Attack} & \textbf{ASR (\%)} & \textbf{Avg Queries} & \textbf{Max Queries} & \textbf{Defense Effective?} \\
\midrule
Square Attack \cite{andriushchenko2020square} & 24.3 ± 2.1 & 2,847 & 5,000 & Yes \\
HopSkipJump \cite{chen2019hopskipjump} & 18.7 ± 1.8 & 1,923 & 3,000 & Yes \\
BPDA+EOT \cite{tramer2020adaptive} & 31.5 ± 2.4 & 38 & 50 & Moderate \\
\midrule
\textbf{Average vs Gradient Attacks} & \textbf{24.8} & \textbf{1,603} & \textbf{-} & \textbf{Consistent} \\
\bottomrule
\end{tabular}}
\end{table}

Our defense demonstrates consistent performance against gradient-free and black-box attacks, with ASR values (18.7–31.5\%) comparable to or lower than gradient-based attacks (31.2–37.2\%). \textbf{Black-box attacks} like Square and HopSkipJump required thousands of queries to reach modest success rates (24.3\% ASR for Square with 2,847 average queries, 18.7\% ASR for HopSkipJump with 1,923 queries), highlighting the increased effort our defense imposes on attackers without gradient access. This query overhead makes practical attacks significantly more expensive and detectable, as production systems typically implement query-based anomaly detection.

For BPDA+EOT, we backpropagate through CABDR with a straight-through estimator and apply EOT over the TRO dropout draws (T=10), achieving 31.5\% ASR with only 38 queries on average. Even this strongest adaptive attempt only moderately succeeds, demonstrating that our defense significantly raises the bar for sophisticated attackers. The consistency of performance across white-box, gradient-free, and black-box attack scenarios indicates genuine robustness rather than gradient masking, addressing concerns raised in recent adversarial robustness literature \cite{tramer2020adaptive}.

\subsection{Functionality-Preserving Attack Assessment}

We evaluate against PE-aware attacks that maintain executable functionality, addressing the critical requirement that adversarial malware remains functional.

\begin{table}[!htb]
\centering
\scriptsize
\caption{Functionality-Preserving PE Attack Results}
\label{tab:functionality}
\setlength{\tabcolsep}{3pt} 
\begin{tabular}{l|c|c|c|c}
\toprule
\textbf{Validation Check} & \textbf{Pass Rate (\%)} & \textbf{ASR (\%)} & \textbf{Func. Preserved} & \textbf{Overall Success} \\
\midrule
PE Structure & 94.2 ± 1.3 & 28.7 ± 2.1 & 27.0 ± 2.0 & 25.5 ± 1.9 \\
Section Integrity & 91.8 ± 1.5 & 29.3 ± 2.3 & 26.9 ± 2.1 & 24.7 ± 1.8 \\
Import Table & 96.1 ± 0.9 & 27.2 ± 1.8 & 26.1 ± 1.7 & 25.1 ± 1.6 \\
Entropy Bounds & 87.3 ± 2.1 & 32.1 ± 2.5 & 28.0 ± 2.3 & 24.4 ± 2.0 \\
\midrule
\textbf{Combined} & \textbf{92.4 ± 1.2} & \textbf{29.3 ± 2.2} & \textbf{27.0 ± 2.0} & \textbf{24.9 ± 1.8} \\
\bottomrule
\end{tabular}
\vspace{-0.1in} 
\end{table}

The functionality-preserving attacks achieve 24.9\% overall success rate, demonstrating that our defense maintains effectiveness even against sophisticated attacks that preserve malware functionality. The high pass rates (87–96\%) for PE validation checks confirm that successful attacks maintain executable structure.

\subsection{Uncertainty Calibration Analysis}

Since TRO relies on uncertainty estimation, we conduct comprehensive calibration analysis using Expected Calibration Error (ECE) and reliability diagrams.

\begin{table}[!t]
\centering
\tiny
\caption{Uncertainty Calibration Results for Different T Values}
\label{tab:calibration}
\setlength{\tabcolsep}{6pt}
\begin{tabular}{c|c|c|c|c}
\toprule
\textbf{T (MC Samples)} & \textbf{ECE} & \textbf{Brier Score} & \textbf{Overhead} & \textbf{Recommended?} \\
\midrule
2 & 0.087 ± 0.008 & 0.162 ± 0.012 & $1.45\times$ & No \\
5 & 0.041 ± 0.005 & 0.089 ± 0.007 & $1.62\times$ & No \\
\textbf{10} & \textbf{0.023 ± 0.003} & \textbf{0.067 ± 0.005} & \textbf{$1.76\times$} & \textbf{Yes} \\
20 & 0.019 ± 0.003 & 0.061 ± 0.004 & $2.31\times$ & No \\
\bottomrule
\end{tabular}
\end{table}
\vspace{0.15in}

Our calibration analysis reveals that T=10 Monte Carlo samples provides the optimal balance between uncertainty quality (ECE=0.023) and computational efficiency ($1.76\times$ overhead). From $T{=}5\to10$, ECE halves (0.041→0.023) while overhead grows modestly ($1.62\times$→$1.76\times$); beyond that, gains taper (Table VIII). This validates our T=10 choice for the TRO mechanism.

\subsection{Threshold Adaptation Analysis}

To evaluate the effectiveness of our dynamic threshold adaptation mechanism, we analyze how the system responds to varying attack intensities and data distributions during evaluation.

The adaptive threshold mechanism demonstrates stable convergence properties across different experimental conditions. During clean sample evaluation, thresholds converge within 2-3 batches to optimal values that maximize throughput while maintaining security. Under adversarial conditions, the system adaptively increases thresholds to trigger more frequent raw feature extraction, successfully maintaining defense effectiveness with only modest computational overhead increases (1.5-1.8× vs baseline 1.76×).

Our ablation study (Table~\ref{tab:ablation}) confirms that static thresholds achieve lower ASR (38.8\%) but at higher computational cost (1.85× overhead), validating the importance of dynamic adaptation for production efficiency. The threshold adaptation provides a practical mechanism for balancing security and performance based on real-time threat assessment, with convergence typically achieved within 2-4 processing batches across diverse evaluation scenarios.

\FloatBarrier

\vspace{0.15in}
\section{Theoretical Analysis}

\subsection{Robustness Guarantees}

We provide theoretical analysis of our defense mechanism's robustness properties under standard threat models.

\textbf{Theorem 1 (Uncertainty-Based Detection)}: \textit{For a classifier $f_\theta$ with Lipschitz constant $L$, if the uncertainty estimate $\mathcal{U}(x)$ satisfies $\mathcal{U}(x + \delta) - \mathcal{U}(x) \geq \gamma \|\delta\|_2$ for adversarial perturbation $\delta$, then the TRO mechanism detects adversarial examples with probability at least $1 - e^{-\gamma^2/2}$.}

\textit{Proof sketch}: The uncertainty increase under adversarial perturbation follows from the Lipschitz property and concentration inequalities for Monte Carlo estimates. The detection probability bound follows from Hoeffding's inequality applied to the uncertainty threshold test.

\textbf{Proposition 1 (Local Non-Sensitivity Region)}: \textit{The CABDR quantization mechanism with bit-depth $b$ creates a local non-sensitivity region of radius $r = 2^{7-b}$ in $\ell_\infty$ norm, where small input perturbations do not affect the quantized representation.}

\textit{Analysis}: The quantization operation $Q_b(x) = \text{round}(x/2^{8-b}) \cdot 2^{8-b}$ creates piecewise constant regions. Within each quantization cell of width $2^{8-b}$, perturbations smaller than $r = 2^{7-b}$ do not change the quantized output, providing local insensitivity to small adversarial modifications.

\subsection{Convergence Analysis}

The dynamic threshold adaptation mechanism exhibits favorable convergence properties.

\textbf{Proposition 2}: \textit{Under mild assumptions on the attack distribution, the adaptive threshold $\tau_t$ converges to a stationary point $\tau^*$ that locally optimizes the efficiency-robustness trade-off with convergence rate $O(1/\sqrt{t})$.}

The convergence analysis follows from standard results in online convex optimization, with the key insight that our objective function is locally convex in the threshold parameter.

\subsection{Computational Complexity}

We analyze the computational complexity of our framework components:

\begin{itemize}
\item \textbf{TRO Mechanism}: $O(T \cdot n)$ for uncertainty estimation with $T$ Monte Carlo samples and batch size $n$
\item \textbf{CABDR}: $O(n \cdot d)$ for quantization of $n$ samples with $d$ features
\item \textbf{Threshold Adaptation}: $O(1)$ amortized cost per batch
\item \textbf{Overall}: $O(n \cdot (T + d))$ per batch, linear in batch size
\end{itemize}
The linear scaling with batch size explains the excellent throughput characteristics observed in experiments.

\section{Discussion}

\subsection{Production Deployment Considerations}

Our framework addresses several critical challenges for production deployment:

\textbf{Resource Allocation}: The adaptive nature of our defense allows dynamic resource allocation based on threat levels. During periods of low attack activity, the system automatically reduces defensive overhead, freeing resources for other tasks.

\textbf{Integration with Existing Systems}: Our framework can be integrated as a preprocessing layer with existing malware detection systems, requiring minimal changes to production pipelines. The modular design allows selective deployment of components based on specific requirements.

\textbf{Monitoring and Maintenance}: The framework provides interpretable metrics (uncertainty levels, threshold values) that facilitate monitoring and debugging. Administrators can track these metrics to identify potential attacks or system anomalies.

\subsection{Comparison with Commercial Solutions}

To contextualize our results, we compare with commercial antivirus solutions. Industry reports indicate that commercial AV products typically process samples at 10–50 ms per file with basic heuristics \cite{harang2020sorel}. Our framework's 76 ms P99 latency represents a modest increase of 26–66 ms over conventional tools, which we argue is acceptable given the substantial security benefits:

\textbf{Latency Justification}: The 76 ms P99 latency is well within acceptable bounds for most deployment scenarios. Email gateways and web proxies targeting sub-100ms response times can comfortably accommodate our framework. For batch analysis (e.g., nightly scans, forensic investigation), latency is not a constraint. Critically, commercial tools offering no adversarial robustness are vulnerable to evasion attacks that our framework mitigates. The marginal latency increase (26–66 ms) prevents potentially catastrophic security breaches that could cost organizations millions in incident response and data recovery.

\textbf{Cost-Benefit Analysis}: While traditional AV solutions offer lower latency, they provide no protection against adversarial malware that can bypass detection with 72\% success rate (Table~\ref{tab:main_results}, Baseline ASR). Our framework reduces this to 34\%, representing a 52\% improvement in robustness for a 52–152\% increase in latency. For high-value targets facing advanced persistent threats (APTs), this trade-off strongly favors robustness over marginal latency gains.

\subsection{Malware Evolution and Concept Drift}

The malware threat landscape evolves continuously, with new malware families emerging daily and existing families adapting evasion techniques. Our framework addresses this challenge through several mechanisms:

\textbf{Temporal Robustness}: Our evaluation uses a temporal split with training samples from 2017–2018 and test samples from 2019, simulating one year of concept drift. The maintained performance (91.0\% clean accuracy, 95.9\% AUC) demonstrates resilience to natural malware evolution over this timeframe.

\textbf{Adaptive Defense Mechanisms}: The TRO mechanism's reliance on uncertainty estimation rather than specific feature patterns provides inherent robustness to concept drift. As new malware variants emerge, the system's uncertainty naturally increases, triggering more thorough raw feature analysis. This self-adjusting behavior allows the framework to maintain effectiveness without requiring immediate retraining.

\textbf{Continuous Learning}: While not implemented in our current evaluation, the framework's modular design supports online learning and periodic model updates. Organizations can deploy continuous monitoring to track performance metrics and trigger retraining when detection rates decline. The efficiency of our approach ($1.76\times$ overhead) leaves computational headroom for periodic model updates without service interruption.

\textbf{Adversarial Evolution}: Beyond natural malware evolution, adversaries may adapt attack strategies specifically targeting our defense mechanisms. Our evaluation against adaptive attacks (BPDA+EOT) demonstrates robustness to attackers with full knowledge of our defense. However, long-term deployment will require monitoring for novel attack patterns and iterative defense improvements—a challenge shared by all security systems.

\subsection{Generalizability to Other Big Data Security Domains}

While our implementation focuses on Windows PE malware detection, the core principles of our framework—efficiency-optimized adversarial defense through confidence-adaptive mechanisms—generalize to other big data security applications:

\textbf{Network Intrusion Detection}: The TRO+CABDR framework can be adapted for network flow analysis by replacing PE-specific raw features with deep packet inspection. The confidence-adaptive quantization principle applies directly to numerical flow features (packet sizes, timing statistics), while the trust-based override mechanism can trigger deeper analysis for suspicious traffic patterns. The efficiency benefits are particularly valuable given the high-volume nature of network traffic (often exceeding 1M flows/second in enterprise environments).

\textbf{Log Anomaly Detection}: Security information and event management (SIEM) systems processing millions of log entries daily face similar efficiency-robustness trade-offs. Our adaptive bit-depth reduction can be applied to numerical log features, while uncertainty-based selective analysis can trigger detailed investigation of suspicious events. The framework's ability to maintain high throughput (1.26M samples/sec) aligns well with big data log processing requirements.

\textbf{Mobile Application Security}: Android malware detection shares structural similarities with PE analysis, using static features extracted from APK files. The raw feature extraction component can be adapted to use Android-specific features (permissions, API calls, DEX bytecode statistics), while the core defense mechanisms remain applicable. The modest computational overhead ($1.76\times$) is acceptable for mobile security scanning services processing millions of app submissions.

\textbf{Cloud Workload Security}: Cloud platforms executing thousands of serverless functions or containers per second require efficient security scanning. Our framework's confidence-based selective analysis aligns with cloud security models where most workloads are benign and only suspicious instances warrant detailed investigation.

\textbf{Implementation Considerations}: Adapting our framework to these domains requires: (1) identifying domain-appropriate raw features that are harder to manipulate, (2) training domain-specific uncertainty estimators, and (3) tuning confidence thresholds based on domain-specific false positive/false negative cost ratios. The core algorithmic principles and theoretical guarantees transfer directly, though domain-specific engineering is necessary for optimal performance.

\subsection{Limitations and Future Work}

While our framework demonstrates strong performance, several limitations warrant discussion:

\textbf{Feature Extraction Scope}: Our throughput measurements focus on inference with pre-extracted EMBER features. In end-to-end deployments, organizations typically cache extracted features or parallelize parsing to maintain performance.

\textbf{Windows PE Focus}: Our implementation targets Windows PE files. Extension to other platforms (Linux ELF, Android APK) requires adapting the raw feature extraction component.

\textbf{Concept Drift}: While our temporal split evaluation provides some evidence of robustness to drift, long-term deployment studies are needed to fully understand performance degradation over time.

\textbf{Future Directions}:
\begin{itemize}
\item Extending the framework to support multi-platform malware detection
\item Investigating federated learning approaches for privacy-preserving updates
\item Exploring hardware acceleration opportunities for further efficiency gains
\item Developing certified robustness guarantees for the complete system
\end{itemize}

\subsection{Broader Impact}

The deployment of effective adversarial defenses has important implications for cybersecurity:

\textbf{Raising the Bar for Attackers}: By making adversarial attacks more expensive and less reliable, our framework increases the cost of successful attacks, potentially deterring less sophisticated adversaries.

\textbf{Enabling Robust ML in Security}: Our work demonstrates that machine learning can be deployed robustly in security-critical contexts, encouraging broader adoption of ML-based security solutions.

\textbf{Economic Considerations}: The efficiency focus of our work makes adversarial defenses economically viable for a broader range of organizations, democratizing access to robust security technologies.

\section{Conclusion}

This paper presented a novel framework for adversarial malware defense that explicitly optimizes the trade-off between robustness and computational efficiency. Through the combination of Trust-Raw Override (TRO) and Confidence-Adaptive Bit-Depth Reduction (CABDR), our approach achieves practical adversarial robustness with only $1.76\times$ computational overhead—a substantial improvement over existing methods requiring $4$–$22\times$ overhead.

Our comprehensive evaluation on the EMBER v2 dataset demonstrates that the framework:
\begin{itemize}
\item Maintains 91\% clean accuracy while reducing attack success rates to 31–37\%
\item Achieves throughput of 1.26 million samples per second, exceeding production requirements by $1.26\times$
\item Provides 82\% cost savings compared to state-of-the-art robust defenses
\item Meets latency requirements for interactive applications with 76 ms P99 latency
\end{itemize}

The success of our approach suggests that practical adversarial robustness requires rethinking traditional optimization objectives. Rather than pursuing maximum robustness without regard to computational costs, we must design defenses that explicitly balance security and efficiency. This paradigm shift is essential for translating academic advances in adversarial robustness into production-ready security solutions.

Our work opens several avenues for future research, including extension to other malware types, integration with threat intelligence systems, and development of certified robustness guarantees. As adversarial threats continue to evolve, the ability to deploy robust defenses efficiently will become increasingly critical for maintaining effective cybersecurity.

\section*{Reproducibility}

To facilitate reproducibility, we provide comprehensive implementation details and experimental configurations. All experiments use PyTorch 2.0.1 with CUDA 11.8, conducted on NVIDIA H100 GPUs with 3rd Gen AMD EPYC processors. We use mixed precision (FP16/FP32) for optimal throughput. Our evaluation protocol includes 5 independent runs with different random seeds (42, 123, 456, 789, 1024) for statistical significance. Complete source code, model configurations, and training scripts are publicly available at: \url{https://github.com/akchaud5/Adversarial-ML-for-Malware-Detection-IEEE-BigData-2025}. Hyperparameters: learning rate $2 \times 10^{-4}$, batch size 256 for training, dropout 0.2, uncertainty thresholds $\tau_{low} = 0.1$, $\tau_{high} = 0.3$, Monte Carlo samples T=10.

\bibliographystyle{IEEEtran}

\end{document}